\documentclass[12pt]{article}
\usepackage{graphicx}
\usepackage{subfigure}
\usepackage[dvips]{color}

\input epsf.tex

\newcommand{\beq}{\begin{equation}}
\newcommand{\eeq}{\end{equation}}
\newcommand{\be}{\begin{equation}}
\newcommand{\ee}{\end{equation}}
\newcommand{\bea}{\begin{eqnarray}}
\newcommand{\eea}{\end{eqnarray}}

\makeatletter
\renewcommand{\theequation}{\thesection.\arabic{equation}}
\@addtoreset{equation}{section}
\makeatother

\def\href#1#2{#2}

\begin{document}

\baselineskip=15.5pt
\pagestyle{plain}
\setcounter{page}{1}

\begin{titlepage}
\begin{flushleft}
       \hfill                       FIT HE - 15-02 \\
       \hfill                       
\end{flushleft}

\begin{center}
  {\huge Entanglement temperature for the excitation of SYM theory   \\ 
   \vspace*{2mm}
in (De)confinement phase \vspace*{2mm}
}
\end{center}

\begin{center}

\vspace*{2mm}
{\large Kazuo Ghoroku${}^{\dagger}$\footnote[1]{\tt gouroku@dontaku.fit.ac.jp},
${}^{}$Masafumi Ishihara${}^{\ddagger}$\footnote[2]{\tt masafumi@wpi-aimr.tohoku.ac.jp},
}\\

%

\vspace*{2mm}
{${}^{\dagger}$Fukuoka Institute of Technology, Wajiro, 
Higashi-ku} \\
{
Fukuoka 811-0295, Japan\\}
{
${}^{\ddagger}$WPI-Advanced Institute for Materials Research (WPI-AIMR),}\\
{
Tohoku University, Sendai 980-8577, Japan\\}

\vspace*{3mm}
\end{center}

\begin{center}
{\large Abstract}
\end{center}
We study the holographic supersymmetric Yang-Mills (SYM) theory, which is 
living in a hyperbolic space, in terms of the entanglement entropy.
The theory contains a parameter $C$ corresponding
to the excitation of the SYM theory, and it controls the dynamical properties of the theory. 
The entanglement temperature, $T_{ent}$, is obtained by imposing the thermodynamic law 
for the relative entanglement entropy and the energy density of the excitation. 
This temperature is available at any value of the parameter $C$ even in the region where
the Hawking temperature disappears.
With this new temperature, the dynamical
properties of the excited SYM theory are examined in terms of the thermodynamic law. 
We could find the signatures of phase transitions of the theory.

\vfill
\begin{flushleft}

\end{flushleft}
\end{titlepage}
\newpage

\vspace{1cm}
\section{Introduction}
The 
holographic approach is a
powerful method to study the non-perturbative properties of 
the strong coupling gauge theories \cite{ads1,ads2,ads3}. 
In this context,
various attempts have been performed to study the properties of the supersymmetric 
Yang Mills (SYM) theory in the confinement phase.
Recently, the quantum information of strong coupling theory has been studied through
the holographic entanglement entropy ($S_{EE}$), which
is very useful to investigate the theory from the thermodynamic viewpoint by supposing
the themodynamic law shown at high temperature \cite{RT}-\cite{KKM}.

As shown in \cite{RT,RT2}, $S_{EE}$ is obtained by separating the space to two
regions $A$ and its complement $\bar{A}$ as follows 
\begin{equation}\label{see}
S_{EE}=\frac{Area (\gamma_A)}{4G_N^{(5)}} ,
\end{equation}
where $\gamma_A$ denotes the minimal surface whose boundary is defined by $\partial A$ and the surface is extended into the bulk. 
$G_N^{(5)}=G_N^{(10)}/(\pi^3R^5) $ denotes the 5D Newton constant reduced from the 10D one $G_N^{(10)}$.
{The area is given as
\beq
 Area (\gamma_A)\equiv S_{\rm Area}(G^{(0)},X^{ext})=\int_{\gamma_A} d^d\xi\sqrt{g}\, ,
\eeq
where the induced metric $g_{ab}$ on $\gamma_A$ are defined as}
\beq
 g=det(g_{ab})\, , \quad g_{ab}=G_{MN}{\partial X^M\over\partial\xi^a}
       {\partial X^N\over\partial\xi^b}\, .
\eeq
The minimal surface $\gamma_A$ is expressed by the profile $X^{ext}$, which is 
embedded in the bulk 
background defined by $G^{(0)}$. This formulation has been extended to 
{the non-conformal case}
in terms of the string frame metric by including non-trivial dilaton \cite{KKM}. 
{Here, we consider the case of the trivial dilaton and the 5D compact space of $S^5$.
Therefore, the above formula is enough.}

{At high temperature in the de-confinement phase,}
it is well known that the entanglement entropy obtained as above approaches to
the thermal entropy, which satisfies the Bekenstein-Hawking relation, in the limit
of large area for the considering system. This fact is convinced
in the high temperature SYM theory, which is dual to the
{AdS$_5$-Schwarzschild gravity}, and the temperature is defined by the Hawking temperature in this case.

On the other hand, in general,
the temperature can not be defined as the Hawking temperature in the 
confinement phase. 
$S_{EE}$ is however calculable by using the same formula {with (\ref{see})}. 
{In other words, the above formula {with (\ref{see})} for $S_{EE}$ 
is useful both in the confinement and the de-confinement phases.}
{In order to see the thermodynamical properties clearly,} the calculations are performed for the large 
{area limit of $\gamma_A$ for} the theory which contains
a parameter $C$ corresponding to the excitation of the theory from its vacuum state. 
\footnote{This parameter has been firstly introduced in the scinario of the brane world \cite{BDEL}-\cite{SSM}. However, its role 
in the present holographic case is different from the case of the brane-world.}
The energy
momentum tensor of this theory is described by this parameter. 
{By the holographic renormalization method \cite{KSS}-\cite{FG}, it has been given in \cite{EGR}. 
Using this energy density,} the entanglement temperature ($T_{ent}$) for the excited state 
is calculated according to the method given in \cite{BNTU}. {$T_{ent}$}
can be defined even if the state is in the confinement phase as far as the excitation due to $C$
exists. Therefore the 
properties of the theory can be
investigated thermodynamically by using this new temperature $T_{ent}$ for over all region of the
parameter $C$.

Our purpose is to investigate the dynamical properties of the 
excited SYM theories, which could have a rich
phase structure, 
by using the thermodynamic laws represented by the temperature
$T_{ent}$. The merit of using this temperature is that $T_{ent}$ is available in
all the region of the excitation parameter. 
Through the analysis given here, we could show the 
important signs of the phase transitions, which have been indicated by 
performing the non-thermodynamic holographic 
analysis \cite{EGR,EGR2}. This indicates that the entanglement temperature $T_{ent}$ is useful
to study the dynamics of the excited state thermodynamically
even if the Hawking temperature disappears.

\vspace{.5cm}
The outline of this paper is as follows.
In the next section, how $T_{ent}$ is defined is explained. In the Sec. 3, 
the holographic SYM theory with the parameter
$C$, which is mentioned above, 
is reviewed, and then the energy momentum tensors are given.
In the Sec. 4, the entanglement entropy is calculated in our model
and discussed in many points through an approximate form. In the section 5, 
entanglement temperature $T_{ent}$ is given. Then its meaning and the thermodynamic
investigations for the two phase transitions of the theory are discussed.  
The summary and discussions are given in the final section.

\section{Entanglement Temperature}

The thermodynamic relation of $S_{EE}$ and the energy density of the system
{has} been related by introducing the modular Hamiltonian ($H$)
as \cite{CHM}
\beq\label{first-law}
  \Delta S_{EE}=\Delta H\, , 
\eeq
\beq
     \rho=e^{-H}\, ,
\eeq
where $H$ is defined as above by the density matrix 
$\rho$ which determines the entanglement entropy as $S_{EE}=-Tr \rho\ln \rho$. 

In the above, usually, the
infinitesimal increasing of $S_{EE}$ and $H$ are given as
\beq
  \Delta S_{EE}=S_{EE}(G^{(0)}+\delta G, X^{(0)})-S_{EE}(G^{(0)}, X^{(0)})=\int d^d\xi\sqrt{g}{g}^{ij}
              \delta g_{ij}\, ,
\eeq
for $S_{EE}$, and for the modular hamiltonian,
\beq\label{Modular-H}
       \Delta H=\int d^d\xi\sqrt{g^{(0)}}\beta\delta\langle T_{00}\rangle\, .
\eeq
where $G^{(0)}$ denotes a solution of the (d+1) dimensional
bulk gravity which is dual to the corresponding
d-dimensional field theory, and $X^{(0)}$ represents the profile $X^{ext}$ of the minimal surface
embedded in the bulk determined by $G^{(0)}$. 
The energy density $\langle T_{00}\rangle$ of the boundary theory is obtained according to the holographic method
for a given $G^{(0)}$ \cite{KSS,BFS}.
The metric on the boundary
is denoted by $g^{(0)}(\neq g)$. Do not confuse $g^{(0)}$ with the induced metric $g$ here.
The factor $\beta$ is introduced as the temperature $\beta=1/T$.
However, in this formulation, it may depend on the coordinates on the boundary and the shape of $\partial A$ as shown in the case
of the CFT.

\vspace{.2cm}
Actually, in the case that $G^{(0)}$ is given by
AdS$_5$, the temperature $1/\beta$ is
obtained as follows according to \cite{WKZV}.
The modular hamiltonian for a ball-shaped region 
with radius $l$ in the Minkowski space is given as \cite{CHM}
\beq\label{CFT}
    H=2\pi\int d^dx {l^2-r_d^2\over 2l}T^{00}(x)\, .
\eeq
{where $r_d=\sqrt{(x^1)^2+\cdots(x^{d})^2}$.}
This implies the temperature defined above as 
\beq
 {\beta}\equiv{1\over T_{ent}}=2\pi{l^2-r_d^2\over 2l}
\eeq
for this CFT case. {Both} $T_{ent}$ and $T^{00}$ are dependent on the coordinate 
on the boundary. Then it is difficult to imagine a thermodynamic picture for the
deviation $\delta g_{\mu\nu}$ which is in general a complicated function of coordinates.

We notice that
the above deviations of $S_{EE}$ and $H$ are obtained in the linear order of
$\delta G$. In this case, it has been shown that the 
relation (\ref{first-law}) is always satisfied when $\delta G$ satisfies the linearized 
5D Einstein equation under the background $G^{(0)}$ \cite{LMR}. 

\vspace{.3cm}
On the other hand, consider a global excitation 
as studied in \cite{BNTU}, 
\beq\label{deltaT1}
\langle T_{00}\rangle=mR^3/(4\pi G_N^{(5)})\, ,
\eeq
where $m$ denotes a parameter corresponding to the excitation in the vacuum 4D
Minkowski space-time. In this case, the bulk metric near the boundary ($z\sim 0$) is given as
\beq
  ds^2=\left({R\over z}\right)^2\left(-{1\over f(z)}dt^2+f(z)dz^2+\sum_{i=1}^{3}d{x^i}^2\right)\, ,
  \quad f(z)=1+mz^4+\cdots
\eeq 
Then consider the deviations of $S_{EE}$ and $H$ for small $m$ within the linear order
of $m$ (not for the metric deviation $\delta G$). In this case, we obtain
\beq\label{temp-0}
  \Delta S_{EE}=\partial_m S_{EE}(G^{(0)}(m), X^{(0)})\bigg|_{m=0}=\beta\int dx^d\sqrt{g^{(0)}}
          \partial_m\langle T_{00}\rangle\bigg|_{m=0}\, ,
\eeq
where $\beta=1/T_{ent}$ is written outside of the integration by assuming it as a constant. 
This assumption seems to be reasonable since
the deviation of (\ref{deltaT1}) is independent of the boundary coordinates. This implies that
the excitation of the system can be observed uniformly through the global temperature 
$T_{ent}$ which is independent of the coordinate.
{We should say however that the entanglement temperature $\beta$ would 
generally depend on the choice of the entanglement region. So the above setting of the equation
(\ref{temp-0}) would be restricted to some special cases of the exitation.}

Within the linear approximation for the parameter $m$, it is possible to estimate $T_{ent}$.
For the case of a ball-shaped region with radius $l$ of the Minkowski space, it is given
for small $l$ as follows {\cite{BNTU},}
\beq
 T_{ent}={5\over 2\pi}{1\over l}\, .
\eeq
For more general cases of global $\langle T_{\mu\nu}\rangle$, the similar evaluation of 
$T_{ent}$ are obtained for small $l$ \cite{BNTU,AAN,Pang}.

\vspace{.2cm}

For large $l$, however, it is necessary to give the metric form at large $z$ up to the deep infrared region. Furthermore,
we need to obtain the profile function of the minimal surface in the infrared region. 
Up to now, the research in this
direction is few.  
Our main purpose is to extend this approach to the infrared region or to the large size area.
We define the temperature $T_{ent}$ at any value of the
parameter $m$ which expresses the excitation of the system. 
\footnote{We notice that the parameter $m$ is used here as a symbolic quantity of the
excitation.}
In the above CFT example, the excitation is seen from the vacuum, $m=0$, to
the excited state with small $m$. So $T_{ent}$ can be defined at $m=0$ as a limit of $m \to 0$. 
In our approach, on the other hand,
we could define the temperature $T_{ent}(m)$ at any $m$ 
by comparing $S_{EE}(m)$ and $S_{EE}(m+\delta m)$. As explained below, this is possible
since we have a holographic solution in which the parameter $m$ is arbitrary. 
Thus we can study the thermal properties of the 
excited state at any value of $m$.

\vspace{.2cm}
{The} definition of the entanglement temperature
$T_{ent}(\alpha)$ is performed 
by using the relative entropy and the thermodynamic first law as follows. 
Here the parameter $m$ is generalized to $\alpha$, and then the energy-momentum of
the excitation is denoted as
$\langle T_{\mu\nu}(\alpha)\rangle$.
The details of this formulation are seen in \cite{CHM,LMR} for the case of CFT.
The relative entropy $S_{EE}(\rho_1|\rho_0)$ is related to $\Delta H$ and $\Delta S_{EE}$
as follows \cite{BCHM,BNTU},
\beq
  S(\rho_1|\rho_0)=\Delta H-\Delta S \geq 0
\eeq
for two density matrices $\rho_1$ and $\rho_0$. 
Consider the case that
the density matrices introduced above are characterized by the parameters $\alpha$
as follows,
\beq\label{alpha}
 \rho_1=\rho(\alpha_1)\,  , \quad \rho_0=\rho(\alpha_0)\, , 
\eeq
where we suppose
\beq
      \alpha_1=\alpha_0+\delta\alpha\, .
\eeq
In the case of infinitesimal small $\delta\alpha$, 
we find the following relation 
\beq
  \Delta S_{EE}=\partial_{\alpha}S_{EE}=\Delta H
\eeq
at $\alpha=\alpha_0$, namely in the limit of $\delta\alpha=0$. 
Further, supposing that the parameter $\alpha$ is global, we can set 
the following relation, 
\beq\label{ent-T}
  \Delta H=\int d^d\xi\sqrt{g^{(0)}}\beta\delta\langle T_{00}\rangle\
                 =\beta(\alpha)\int d^d\xi\sqrt{g^{(0)}}\partial_{\alpha}\langle T_{00}\rangle\bigg|_{\alpha=\alpha_0}\
\eeq
which has the same form with (\ref{Modular-H}). 
We notice that the second equation of (\ref{ent-T}) is obtained supposing the uniformity of $\beta$
according to the above equation (\ref{temp-0}). 
This setting seems
to be consistent with our analysis given here.
\footnote{For our present case, the ingredients to
realize the Eq. (\ref{ent-T}) are considered as the followings. 
The excitation parameter is a global constant and then the
energy density of the excitation is also global. Furthermore, a large volume limit of entanglement region is considered
here.}

Finally, we arrive at the following formula
for the entanglement temperature,
\beq\label{global-T0}
  T_{ent}^{\alpha}(\alpha_0)={\int d^d\xi\sqrt{g^{(0)}}{\partial\langle T_{00}\rangle\over\partial \alpha}\over
                   {\partial S\over\partial \alpha}}\bigg|_{\alpha=\alpha_0}\, .
\eeq
As mentioned above, notice that
this temperature does not depend on the coordinate but does on the parameter, $\alpha_0$, which determines
the excited state of the theory. {Furthermore, the above formula (\ref{global-T0}) is available at any $\alpha_0$. 
So it is possible to obtain $T_{\rm ent}(\alpha)$ as a function of $\alpha$.}

\section{Gravity dual of excited SYM theory}

\subsection{Model} 

The holographic dual to the large $N$ gauge theory embedded in FRW space-time with two parameters 
is given as the following form of 10D metric \cite{EGR}
\beq\label{10dmetric-2}
ds^2_{10}={r^2 \over R^2}\left(-\bar{n}^2dt^2+\bar{A}^2a_0^2(t)\gamma_{ij}(x)dx^idx^j\right)+
\frac{R^2}{r^2} dr^2 +R^2d\Omega_5^2 \ . 
\eeq
\beq\label{AdS4-30} 
    \gamma_{ij}(x)=\delta_{ij}\left( 1+k{\bar{r}^2\over 4\bar{r_0}^2} \right)^{-2}\, , \quad 
    \bar{r}^2=\sum_{i=1}^3 (x^i)^2\, ,
\eeq
where $k=\pm 1,$ or $0$. {The scale parameter of three space
is denoted by $\bar{r_0}$. }
The solution is obtained within the 10D supergravity
of type IIB theory as follows 
\bea
 \bar{A}&=&\left(\left(1+\left({r_0\over r}\right)^2\right)^2+\left({b_0\over r}\right)^{4}\right)^{1/2}\, , \label{sol-10-1} \\
\bar{n}&=&{\left(1+\left({r_0\over r}\right)^2\right)^2
- \left({b_0\over r}\right)^{4}\over 
       \sqrt{\bar{A}}}\, , \label{sol-11-1}
\eea
\beq
   r_0={R^2\over 2}\sqrt{|\lambda|}\, ,   \quad 
              b_0={R\over a_0}\left({CR^2\over 4} \right)^{1/4}\,  . \label{sol-12-1}
\eeq
Two dimensionful parameters, $\lambda$ and $C$ are introduced in solving the equation of motion.
The parameter $C$, which is called as the "dark radiation'', is introduced as an integration constant.  
On the other hand, the dark energy $\lambda$, which corresponds to the 4D cosmological 
constant, is introduced by the following relation,
\beq
  \left({\dot{a}_0\over a_0}\right)^2+{k\over a_0^2}=\lambda\, . \label{bc-3-1}
\eeq
in solving the bulk Einstein equation. We should notice that the above equation is not
introduced to solve the 4D Einstein equations with the 4D cosmological constant.
Although the value of $\lambda$ is arbitrary, the above bulk solution is considered for 
the negative value of $\lambda(=-|\lambda|<0)$ in order to study the parameter region
where the phase transition occurs. 

Here we comment on the time dependence of the scale factor $a_0(t)$. Its time
dependent form is given by solving (\ref{bc-3-1}).
In our analysis, we consider the case of very small time derivative of $a_0(t)$
for simplicity. For the sake
of the justification of our assumption for $a_0(t)$, we should say that the solution
of constant $a_0$ is supposed here as $a_0=1/\sqrt{|\lambda|}$, which
is allowed for negative constant $\lambda$ when we take $k=-1$.

Then the theory is considered in the 3D hyperbolic space.
In this case, for a fixed $|\lambda|$, the theory shows two phase transitions with increasing $C$
\cite{GNI14,GITT}.
At small $C$, the theory is in the confinement and broken chiral symmetry phase ((A)).
With increasing $C$, de-confinement and broken chiral symmetry phase ((B)) appears. Finally 
de-confinement and restoration of the chiral symmetry phase ((C)) is realized. 
In terms of $(r_0,b_0)$, these phases are assigned as
\beq
   {\rm (A)~} b_0<r_0\, , \quad {\rm (B)~} r_0<b_0<1.31r_0\, \quad
   , \quad {\rm (C)~} 1.31r_0<b_0\, .
\eeq
The transition from (B) to (C) has been discussed in \cite{GITT}.
Then we expect that the dynamical properties of each phase of the theory would be 
observed also as thermodynamic properties in terms of the entanglement temperature
which is defined by using the excitation corresponding to $C$.


\subsection{Energy momentum tensor and meaning of $C$}

For the later convenience, we show the 4D stress tensor of the dual field theory for the present model. It has
been given in \cite{EGR}, 
according to the holographic renormalization method \cite{BFS,KSS}
based on the Fefferman-Graham framework \cite{KSS,BFS,FG}. 
We obtaine the following results,
\beq
 \langle T_{\mu\nu}\rangle=\langle \tilde{T}_{\mu\nu}^{(0)}\rangle+
{4R^3\over 16\pi G_N^{(5)}}\left\{{3\lambda^2\over 16}\left(1,~-{g}_{(0)ij}\right)\right\}\, .
\label{Feff6}
\eeq
\beq
   \langle \tilde{T}_{\mu\nu}^{(0)}\rangle={4R^3\over 16\pi G_N^{(5)}}
{\tilde{c}_0\over R^4}(3,~{g}_{(0)ij})\, , \quad
\eeq
where $G_N^{(5)}=8\pi^3{\alpha'}^4g_s/R^5$, $R^4=4\pi N{\alpha'}^2g_s$
and ${g}_{(0)ij}$ denotes the three dimensional metric on the boundary.
The first part, $\langle \tilde{T}_{\mu\nu}^{(0)}\rangle$, comes from 
the conformal YM fields given in {\cite{GN13}.}
The second term corresponding to the loop corrections of the YM fields 
leads to the conformal anomaly as follows
\beq
 \langle T_{\mu}^{\mu}\rangle=-{3\lambda^2 \over 8\pi^2}N^2\, .\label{anomaly2}\,
\eeq

\vspace{.3cm}

Next, we notice the holographic meaning of the "dark radiation" $C$. The situation 
is different from the case of the brane cosmology. Its meaning is clearly understood
at $|\lambda|=0$ or $r_0=0$, where we find the AdS-Schwarzschild metric.
Then the Hawking temperature is found as
\beq\label{H-T}
   T_H\bigg |_{r_0=0}\equiv T_H^{(0)}= {\sqrt{2}b_0\over \pi R^2}\, .
\eeq
The energy density is given as
\beq\label{energy-density}
  \rho=\langle T_{00}\rangle={3N^2\over 8\pi^2}{T_H^{(0)}}^4\, ,
\eeq
which represents the Stefan-Boltzmann law of the radiation.
This implies that $C$ corresponds to the thermal radiation of SYM fields 
in the 4D Minkowski space-time. 

\vspace{.3cm}  
In the present case, however, we are considering the SYM theory in the 4D curved space-time
which is characterized by $\lambda$. Then the relations
(\ref{H-T}) and (\ref{energy-density}) are modified by the curvature. As a result,
the meaning of $C$ might be changed.
In the deconfinement phases, (B) and (C) with $|\lambda|\neq 0$, $C$ still represents the thermal 
SYM fields. However the formula (\ref{H-T}) is modified as
\beq\label{HawkingT}
    T_H= {\sqrt{2}b_0\over \pi R^2}\sqrt{1-(r_0/b_0)^2}\, .
\eeq
In the confinement phase (A), on the other hand, the temperature 
$T_H$ disappears even if
$C\neq 0$. Then we expect that the glueball like matter 
may be represented by $C$. 
We could find a hint for this expectation by studying the entanglement entropy
in these different phases by changing the value of $C$.

\vspace{.5cm}
\section{Entanglement Entropy}

In order to estimate $T_{ent}$, we firstly examine the entanglement entropy for a sphere
with radius $p_0$.

\vspace{.3cm}
\subsection{Minimal surface }

\begin{figure}[htbp]
\vspace{.3cm}
\begin{center}
\includegraphics[width=7.0cm,height=7cm]{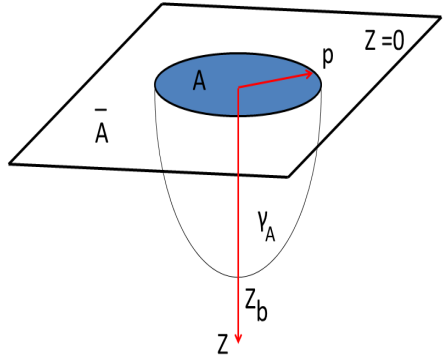}
\caption{ The minimal surface $\gamma_A$ is shown schamtically.   \label{minimal-s}}
\end{center}
\end{figure}

For the case of the present holographic theory, 
from (\ref{10dmetric-2}),  the spatial part of the bulk metric is rewritten as 
\begin{equation} \label{10dspace-r}
ds_{space}^2=\frac{1}{R^2}\left(r^2+2r_0^2+\frac{r_c^4}{r^2}\right)ds^2_{FRW_3}+\frac{R^2}{r^2}dr^2+R^2d\Omega_5^2 ,
\end{equation}
where
\begin{equation}\label{gp-0}
ds^2_{FRW_3}=a_0^2(t)\gamma^2(dp^2+p^2d\Omega_2^2) ,
\end{equation}
\begin{equation}\label{gp-1}
p=\frac{\bar{r}}{\bar{r_0}},\quad\gamma=1/(1-p^2/4) ,
\end{equation}\label
and $r_c$ is defined as
\begin{equation}
r_c\equiv (b_0^4+r_0^4)^{1/4}\, .
\end{equation}
{We used this coordinate since it is useful to study the large scale region by the finite
radial coordinate $p$. In the appendix A, we give a small comment of this coordinate.}

As shown below,
the point $r=r_c$ is called as the domain wall
since the profile of the minimal surface cannot penetrate this point to the infrared region. Namely the solution is restricted to the
region $r_c<r<\infty$.

Here, for the convenience, we
change the variable $r$ to $z$ as
\begin{equation} \label{rcz}
z=r_c^2/r ,
\end{equation}
so it will be restricted to $0<z<r_c$. In this case,
the spatial part of the bulk metric (\ref{10dspace-r}) is rewritten as   
\begin{equation} \label{10dspace-z}
ds_{space}^2=\frac{1}{R^2}\left(z^2+2r_0^2+\frac{r_c^4}{z^2}\right)ds^2_{FRW_3}+\frac{R^2}{z^2}dz^2+R^2d\Omega_5^2 .
\end{equation}

\vspace{.3cm}

We consider  an entangling surface at $z=0$ as a  ball with the radius $p_0$.
Here the profile of the minimal surface in the bulk is set by $p(z)$ which is determined
later.
Then the area of the minimal surface with this boundary, as shown in the Fig. \ref{minimal-s}, is given by \cite{GNI14}
\begin{equation} \label{sarea}
\frac{S_{Area}}{4\pi}=\int_0^{z(p=0)} dz \mathcal{L}(z) , 
\end{equation}
where
\begin{equation} \label{lz}
\mathcal{L}(z)\equiv p(z)^2B\sqrt{Bp'(z)^2+\frac{R^2}{z^2}} ,
\end{equation}
and 
\begin{equation}\label{B}
B\equiv \frac{a_0^2\gamma^2}{R^2}\left(z^2+\frac{r_c^4}{z^2}+2r_0^2\right) .
\end{equation}
\begin{figure}[htbp]
\vspace{.3cm}
\begin{center}
\includegraphics[width=7.0cm,height=7cm]{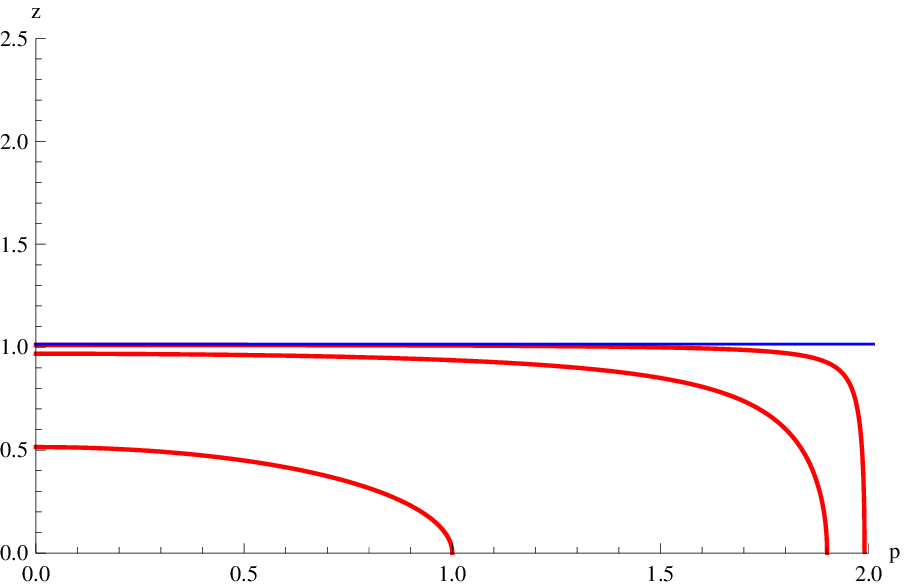}
\includegraphics[width=7.0cm,height=7cm]{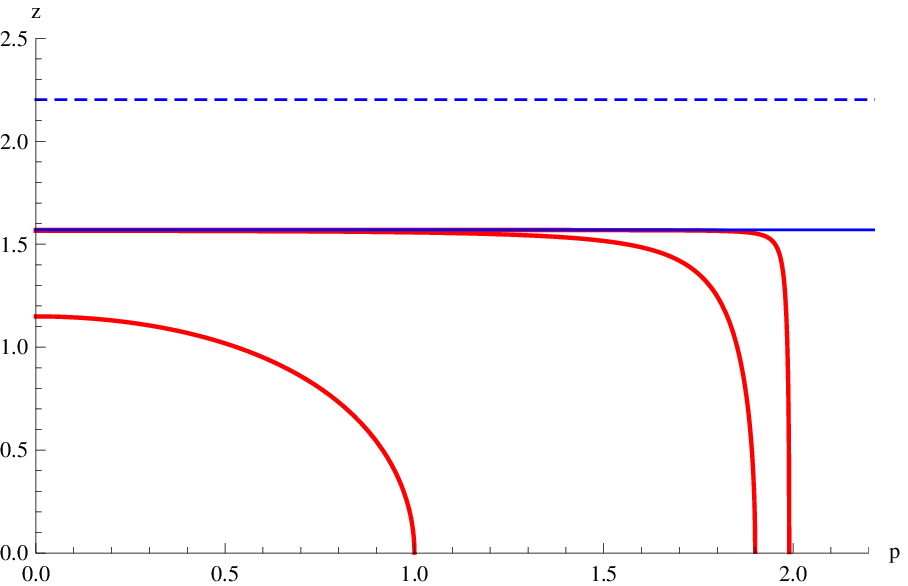}
\caption{{\bf Left phase ;} Embedded solutions for $p(z)$ for $p_0=0.5$, $1.9$ and $1.99$ with $r_0=R=1, a_0=0.5$, $b_0=0.5(<1)$. The blue line is the  domain wall  $r_c=1.02$. {\bf Right phase ;}
 Embedded solutions for $p(z)$ for $p_0=0.5$, $1.9$ and $p_0=1.99$ with $r_0=R=1, a_0=0.5$, $b_0=1.5(>1)$. The blue line is the domain wall $r_c=1.57$ and the dashed blue line is the event horizon $z_H=2.20$.\label{pz10}}
\end{center}
\end{figure}

By solving the variational equation, which is obtained from (\ref{sarea}), we can get the profile  $p(z)$ of the minimal surface. The numerical solutions for confinement phase ($b_0<r_0$) and de-confinement  phase ($r_0\le b_0$)  are shown  in Fig. \ref{pz10} where $p_0$ denotes the ball radius, 
\begin{equation}\label{bound}
 p_0\equiv p(z=0) \leq 2.
\end{equation} 
The upper bound comes from its definition.

\vspace{.2cm}

From these numerical results, we can see that the profile function $p(z)$ approaches to
the rectangle form, namely the bottom line ($z=$const.) and the side lines ($p=$const.)
which are given as 
\beq\label{rectangle}
    z=z_b\, , \quad   p=p_0\, ,
\eeq
respectively, where $p_0$ approaches to the upper limit $p_0=2.0$. 
This behavior 
is 
proved as follows.
The above Eq.  (\ref{sarea}) is rewritten as
\begin{equation} \label{sareaC}
\frac{S_{Area}}{4\pi}=\int_0^{p_0} dp \mathcal{L}(p) , 
\end{equation}
where
\begin{equation} \label{lp}
\mathcal{L}(p)\equiv p^2B\sqrt{B+\frac{R^2 \dot{z}^2(p)}{z^2{(p)}}}\, , \quad \dot{z}(p)={\partial z {(p)}
\over \partial p}\, ,
\end{equation}
and B is the same form with the one given by (\ref{B}).

In this case, the configuration for the minimal surface
is obtained by solving the equation of motion for $z(p)$ which is derived from the
above action as
\beq\label{zp-eq}
  {{\partial\over\partial z(p)}}\left( p^2 B \sqrt{B+\frac{R^2}{z^2}\dot{z}^2(p)} \right)
   -{{\partial\over\partial p}}\left( {p^2 B \frac{R^2}{z^2}\dot{z}(p) \over
             \sqrt{B+\frac{R^2}{z^2}\dot{z}^2(p)} } \right)=0\, .
\eeq
Now, we concentrate on near the top of the minimal surface, namely near $p=0$.
Here, the following relations
\beq\label{conditions}
   \dot{z}(0)=0\, , \quad {\rm and } \quad \ddot{z}(0)<0
\eeq
should be satisfied. From Eq. (\ref{zp-eq}), we find the following equation
\beq
   {a_0^2\gamma^2\over R^2}z_b\left(1-{r_c^4\over z_b^4}\right)
      =\frac{R^2}{3z_b^2}\ddot{z}(0)\,  ,
\eeq
where $\dot{z}(0)=0$ is imposed. Then we find
\beq
    z_b\leq r_c=(b_0^4+r_0^4)^{1/4}
\eeq
to satisfy the second condition of (\ref{conditions}). This implies that the upper bound of
$z$ is given by $r_c$ which is called here as the "domain wall".

\vspace{.2cm}
We notice  the relation of the positions of the domain wall $r_c$ and the horizon $r_H$.
The latter appears only in the de-confinement phase.  By setting as
$z_H\equiv r_c^2/r_H$, we find 
\begin{equation}
z_H^4-r_c^4=2b_0^2r_0^2\left(\frac{z_H}{r_c}\right)^4\ge 0 .
\end{equation}
This implies that the domain wall $r_c$ is  smaller than the horizon $z_H$. Then
the minimal surface could not reach at the horizon even if it appears. This fact
implies that the form of the minimal surface is always connected. This point is
important in this analysis.

\subsection{Entanglement Entropy in the Infrared Limit}
In \cite{KKM}, it has been pointed out that the configuration of the minimal surface changes
from connected one to the disconnected one in the confinement phase. 
In our case, {we find no such topological change of the minimal surface configuration.}
The
minimal surface has the connected form for all region of $p_0$ {though} the theory is in
the confinement phase for $b_0/r_0<1$. This fact is not contradict with the statement of
\cite{KKM} since both the bulk geometry and the shape of the divided region in our case
are different from those studied in \cite{KKM}.

In order to estimate the entanglement entropy by an approximate formula, it is
considered in the infrared limit of $p_0\to 2$. In this limit,
the minimal surface is estimated by substituting the obtained profile function
$z(p)$ into the Eq. (\ref{sareaC}). At the limit of $p_0\to 2$, as shown above, the profile
is approximated by the rectangle form (\ref{rectangle}).  Thus 
$S_{Area}$ of (\ref{sareaC}) can be approximated as
\bea \label{sareaC2}
\frac{S_{Area}}{4\pi}&=&\int_{0, z=z_b, \partial z/\partial p=0}^{p_0} dp p^2B^{3/2}+
                \int_{0, p=p_0, \partial p/\partial z=0}^{z_b}dzp^2B{R^2\over z}\, \\
      &=&\int_{0}^{p_0} dp p^2\left( {a_0^2\gamma^2\over R^2}f(z_b)\right)^{3/2}+
                \int_{0}^{z_b}dzp_0^2{a_0^2\gamma^2(p_0)}g(z)\,           
\eea
where
\bea
  f(z_b)&=&z_b^2+{r_c^4\over z_b^2}+2r_0^2\, , \\
  g(z)&=&z+{r_c^4\over z^3}+2{r_0^2\over z}\, .
\eea
Here we notice that
the first term is dominant for $p_0\to 2$ since it increases with the volume of $A$. On the other
hand, the second term increases with the surface of $A$. We could see that the first term
has its minimum at $z_b=r_c$ from the form of $f(z_b)$ given above. Then we could
understand that the value $r_c$ 
corresponds to the domain wall for the minimal surface.

\vspace{.3cm}
Another point to be noticed is that $r_c$ is larger than the horizon $r_H$
in the de-confinement phase. Then the minimal surface 
bounded at $r=\infty$ could not touch $r_H$ {and} there appears no disconnected 
surface as mentioned above.

\vspace{.3cm}
{We estimate the area of the minimal surface $S_{Area}$ 
by separating into two parts as follows,}
\bea 
\frac{S_{Area}}{4\pi}&=&I_{bottom}+I_{side}, \label{area-1} \\
        I_{bottom}&=&\int_{0}^{p_0} dp p^2\left( {a_0^2\gamma^2\over R^2}f(z_b)\right)^{3/2}\, ,\label{area-2} \\
        I_{side}&=&\int_{0}^{z_b}dzp_0^2{a_0^2\gamma^2(p_0)}g(z)\,   . \label{area-3}
\eea
The two parts, $I_{bottom}$ and $ I_{side}$ are
corresponding to the parts of $z=z_b$ and $p=p_0$ respectively. The first term is given
as
\bea
   I_{bottom}&=&{\bar{V}_{(3)}\over 4\pi R^3}2\sqrt{2}\left(1+h\right)^{3/2}\, , \\
   \bar{V}_{(3)}&=&{\pi R^6\over 2} \int_{0}^{p_0} dp p^2{\gamma^3(p)}\, , \\
        \quad   h&=&\sqrt{1+{b_0^4\over r_0^4}}=\sqrt{1+{4C\over R^2}}\, ,
\eea
where we used 
\beq
   a_0={1\over \sqrt{|\lambda|}}={R^2\over 2r_0}\, .
\eeq
It is noticed that $I_{bottom}$ \footnote{ Notice that ,at the limit of $p_0=2$, the three 
volume is divergent as $$
{2\bar{V}_{(3)}\over \pi R^6}=\int_0^{p_0}\gamma^3p^2dp=\frac{1}{2}a_0^3\left(\frac{4p_0(4+p_0^2)}{(p_0^2-4)^2}+\log\frac{2-p_0}{2+p_0}\right)\, .$$  }
does not contain the parameter $r_0$ or $\lambda$.
Then the first term is expressed by $C$ only. This point is important as seen below.

\vspace{.2cm}
As for the second term $I_{side}$, it can be evaluated by introducing ultraviolet cutoff
$\epsilon$ as
\bea
   I_{side}&=&a_0^2p_0^2\gamma^2(p_0)\left( {z_b^2\over 2}-{r_c^4\over 2z_b^2}
          +2r_0^2\ln {z_b\over r_c}-{\epsilon^2\over 2}+{r_c^4\over 2\epsilon^2}
          -2r_0^2\ln {\epsilon\over r_c}\right).
\eea
Here we take the limit of $\epsilon\to 0$ by subtracting two divergent terms {and}
we get
\beq
  I_{side}=a_0^2p_0^2\gamma^2(p_0)\left( {z_b^2\over 2}-{r_c^4\over 2z_b^2}
          +2r_0^2\ln {z_b\over r_c}\right)+F_s\, .
\eeq
This last term $F_s$ denotes an ambiguity of the subtraction. This is usualy determined 
by an appropriate boundary {conditions} or the renormalization conditions. 

{Our} purpose is to see the change of the entanglement entropy when the excitation $C$
increases, so we take the following boundary {condition,} 
\beq
   \frac{S_{Area}}{4\pi} {\Bigg |}_{z_b=r_c,~C=0} =0\, .
\eeq
Then we find
\beq
  F_s=8{\bar{V}_{(3)}\over 4\pi R^3}\, ,
\eeq
{and} by setting as $z_b=r_c$, $S_{Area}$ is given as
\beq\label{S-finite}
  \frac{S_{Area}}{4\pi} =8{\bar{V}_{(3)}\over 4\pi R^3}\left(
       \left({1+h\over 2}\right)^{3/2}-1\right)\, .
\eeq

As for this result, we {notice} the following points;

\begin{itemize}


\item The result (\ref{S-finite}) indicates that the minimal surface $S_{Area}$ is
independent of $r_0$. This implies that the entanglement temperature is determined
by changing $b_0$ since the entanglement entropy is controlled only by $b_0$.  
The change of $\lambda$ is related to the change of the mass of excited state and
the vacuum energy as seen below.

\item At large $C$ (or equivalently at large $b_0$), {we find,} 
\beq\label{largeb-S}
  S_{EE}={\frac{S_{Area}/V_{(3)}}{4G_N^{(5)}}}={\pi^2\over 2}N^2T^3_H \,.
\eeq
This indicates the entanglement entropy at large scale limit satisfies
the thermodynamic {relation,}  
\beq
   {\partial U\over \partial S_{EE}}=T
\eeq
where $T=T_H$ and 
\beq\label{largeb-U}
       U=\langle T_{00}\rangle = {3\pi^3 R^3\over 16G_N^{(5)} } T_H^4\, .
\eeq
The above formula for $\langle T_{00}\rangle$
is obtained at large $C$ \cite{EGR}.

\item The resultant form of the minimal surface is determined
only by the first term of (\ref{area-1}), which represents the bottom part of the surface.
In other words, the entropy of the excitations due to the dark radiation are given by
the area at the bottom. 

\end{itemize}

\vspace{.1cm}
\subsection{About Logarithmic Divergent term}

{We} comment on the divergent terms in $S_{Area}$. They are found in $I_{side}$ as
\beq\label{div-10}
  I_{side}\bigg|_{div}=a_0^2p_0^2\gamma^2(p_0)\left( {r_c^4\over 2\epsilon^2}
          -2r_0^2\ln {\epsilon\over r_c}\right)\, .
\eeq
However, this is not equivalent to the one given in \cite{GNI14}. The coefficient of the
logarithmic divergent term is slightly different from the above formula (\ref{div-10}).
This point is improved by adding a correction term in $I_{side}$, which is roughly 
approximated. In getting (\ref{div-10}), we have approximated as
\beq\label{p-z}
   p^2{a_0^2\gamma^2(p)}=p_0^2{a_0^2\gamma^2(p_0)}.
\eeq
{Then} the integration with respect to $z$ is performed. This procedure corresponds to
adopt the approximation of $p(z)=p_0$. 
On the other hand, $g(z)$ has a term proportional to $1/z^3$ {and} we must retain
the terms up to $z^2$ in (\ref{p-z}) in order to see the logarithmic divergent terms.

Near $z=0$, the asymptotic solution can be obtained as {\cite{GNI14},}
\beq\label{pzuv}
p=p_0+p_2z^2+p_4z^4+p_{4L}z^4\log z\cdots ,
\eeq
where $p_4$ is an arbitary constant. $p_2$ and $p_{4L}$ are determined as
\begin{equation}\label{p2}
p_2=-\frac{(1-(p_0^2/4)^2)R^3}{2a_0^2p_0r_c^4}\, ,
   \quad p_{4L}=-\frac{\left(1-(p_0^2/4)^2\right)R^8\dot{a}_0^2}{4a_0^4p_0r_c^8} .
\end{equation}

Then, instead of (\ref{area-3}), we obtain
\bea
I_{side-2}&=&\int_{0}^{z_b}dzp_0^2{a_0^2\gamma^2(p_0)}\left(1+s_2z^2\right)g(z)\,  , 
            \label{area-31}\\
           s_2 &=&{2(1+p_0^2/4)\over p_0(1-p_0^2/4)}p_2. 
\eea
In this case we have the divergent term as
\beq\label{div-2}
  I_{side-2}\bigg|_{div}=a_0^2p_0^2\gamma^2(p_0)\left( {r_c^4\over 2\epsilon^2}
          +\left({(1+p_0^2/4)^2\over 2a_0^2p_0^2}R^3-2r_0^2\right)\ln {\epsilon\over r_c}\right)\, .
\eeq
Then we could find the familiar {formula,}
\beq
  {S_{Area}\over 4G_N^{(5)}}=N^2\ln \epsilon +\cdots\, .
\eeq

So we could see the correct form of logarithmic divergence contribution by using the 
higher order term of $p(z)$ with respect to $z$. However, this is useful in the region
of small $z$, and we should be careful about this expansion to large $z\sim z_b$ with large
$b_0$. So hereafter we adopt the formula (\ref{area-3}) in {this} discussion.

\vspace{.5cm}
\section{Excitation and Entanglement Temperature}

\subsection{Entanglement Temperature}

{We} calculate the entanglement temperature for the excitation 
in the SYM theory given
above.  The global temperature $T_{ent}$ is calculated according to the formula 
(\ref{global-T0}) by replacing the parameter $\alpha$ to $b_0$. 
Then the meaning and the role of $T_{ent}$ are investigated for our model in the confinement
phase as well as in the de-confinement phase.
In the latter case, as pointed above,
the entanglement temperature $T_{ent}$ approaches to the Hawking 
temperature $T_{H}$ at large $b_0$ and infrared
limit, namely for the large scale minimal surface 
\cite{BNTU}. We could also see this behavior in our model.

On the other hand,
in the confinement phase,  $T_{H}$ disappears and then
$T_{ent}$ would be used to measure the energy of the system  due to the 
excitation of the SYM fields in the form of the glueballs.
We therefore expect that the mass of the glueball will be related to $T_{ent}$ in the 
confinement phase in some way.

\vspace{.3cm}

In the ultraviolet region, namely
at small $p_0$, we expect the following behavior as given for the de-confining theory
in \cite{BNTU},
\beq
  T_{ent}={c_0\over p_0}
\eeq
where $c_0$ is a calculable number. The reason why this is expected is that
the dynamical properties in the infrared region would not affect on the
quantities at short range physics. 
{In fact} we could see it numerically.
Here our purpose is to
examine the properties at the infrared limit of $p_0\sim 2$ by using
an approximate and simple form of the minimal surface given in our analysis.

\begin{figure}[htbp]
\vspace{.3cm}
\begin{center}
\includegraphics[width=12cm]{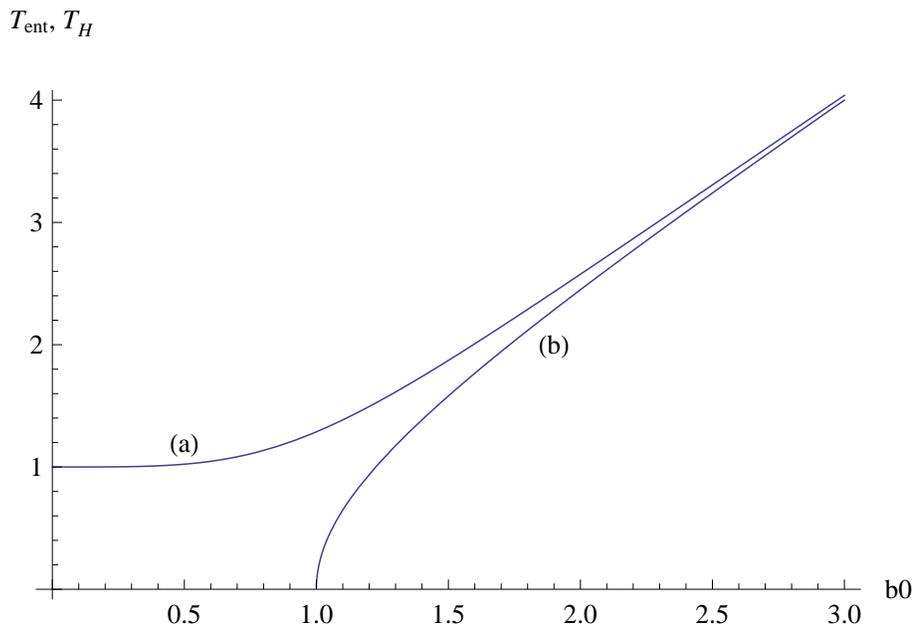}

\caption{(b) $T_H$ and (a) $T_{\rm ent}$ are shown as the function of $b_0/r_0$ for $r_0=1$.
}
\label{Fr}
\end{center}
\end{figure}

\vspace{.3cm}
In our model, there are two parameters, $b_0$ and $r_0$, as the candidates for the above $\alpha$,
which {are} used to define the relative entropy, in (\ref{global-T0}).
Both $b_0$ and $r_0$ may be considered as such parameters. However,
as seen from Eq. (\ref{S-finite}),
the entanglement entropy is expressed as a function of $x^4(=b_0^4/r_0^4)$ which is rewritten as
\beq
   x^4=\left({b_0\over r_0}\right)^4=4{C\over R^2}\, ,
\eeq
where $C$ represents the excitation of the SYM fields. {This means that the entanglement
temperature $T_{ent}$, which should reflect the excitation of the system, is 
determined by the parameter $C$ only. In other words, it is controlled only by the parameter $b_0$. }
Thus, according to (\ref{global-T0}), we have
the entanglement temperature $T_{ent}$ as follows,
\beq\label{global-T}
  T_{ent}^{b_0}(b_0,~r_0)\equiv T_{\rm ent}^{b_0} ={\int d^d\xi\sqrt{g^{(0)}}{\partial\langle T_{00}\rangle\over\partial b_0}\over
                   {\partial S \over\partial b_0}}\, .
\eeq
In the limit of $p_0\to 2.0$, the maximum of $p_0$, we obtain the following result,  
\beq\label{T-entb}
  T_{\rm ent}^{b_0}=
       {\sqrt{2}r_0\over \pi R^2} {h\over \sqrt{1+h}}\, ,
\eeq
where we used (\ref{S-finite}).

\vspace{.3cm}
{In order to make clear the difference between $T_{ent}$ and $T_H$,
we compare them. Here and in the followings, $T_{\rm ent}^{b_0}$ and the Hawking 
temperature $T_H(r_0)$, which is given in (\ref{HawkingT}), are denoted simply 
as $T_{\rm ent}$ and $T_H$ respectively.}
They are shown in the Fig. \ref{Fr}, and it shows $T_H<T_{\rm ent}$ and $T_{\rm ent}$
approaches to $T_H$ from the above at large $b_0$.
The ratio of the two temperatures is expressed as
\beq\label{T-ent}
   T_H/T_{\rm ent}={1\over h}\sqrt{(x^2-1)(1+h)}<1\, .
\eeq
Then, 
at high temperature, we can use {both} temperatures to examine the
thermodynamic properties of the excited system.

On the other hand, while $T_H$ can not be defined in the small $b_0(<r_0)$ region of the 
confinement phase (A), $T_{ent}$ survives in this region and can be
defined in all region where the excitation due to $C$ exists.
So we use $T_{ent}$ instead of $T_H$ 
to see the thermo-dynamical properties in the whole region of $C$. In other words,
it would be possible to extend the thermodynamic viewpoint 
by using the new temperature $T_{ent}$.
From this viewpoint, our model is studied thermodynamically 
in terms of $T_{ent}$ as follows.

\subsection{Thermodynamic properties in terms of $T_{ent}$}

In order to see the thermodynamic properties of the excitation, we consider the 
{quantities given by }the following equations,
\beq
  E(b_0)/{T^4_{\rm ent}}={3\pi^2\over 8}N^2f_U(x)\, , \quad 
   f_U(x)={(1+h)^{2}\over h^4}(h^2-1)\, , \quad h(x)=\sqrt{1+x^4}\, ,
\eeq
\beq
  S_{EE}/{T^3_{\rm ent}}={\pi^2\over 2}N^2f_S(x)\, , \quad 
   f_S(x)={(1+h)^{3/2}\over h^3}\left((1+h)^{3/2}-2\sqrt{2}\right)\, ,
\eeq
where $T_{\rm ent}$ is given by (\ref{T-entb}) and
\beq
  E(b_0)\equiv \langle T_{00}\rangle-\langle T_{00}\rangle\big |_{b_0=0}\, .
\eeq
We investigate the above quantities, $f_U$ and $f_S$, along the value of $b_0$.

\vspace{.3cm}
\noindent{\bf At large $b_0$}; 

We notice that the usual
thermodynamic relations, (\ref{largeb-S}) 
and (\ref{largeb-U}), are obtained at large $b_0$ (or small $r_0$), where we find
$f_U\to 1$, $f_S\to 1$, and $T_{\rm ent}\to T_H$. Then, at large $b_0$, these two 
relations approach to the one obtained at high temperature de-confining phase.
When $b_0$ {decreases,}  $f_U$ and $f_S$ deviate from one. This is the reflection of
the interaction of the dynamical freedom of the excited fields. Then we could see
the dynamical properties of the excited fields through this deviation.

\vspace{.3cm}
\noindent{\bf Near the region of $b_0=0$}; 

In the region of $b_0(<r_0)$, the theory is in the confinement phase {and}  the excitation
is expected to be the color singlet, namely the glueball.
We consider the limit of $b_0=0$, where we have the lowest entanglement 
temperature, which is given as {follows,} 
\beq
   T_{ent}^{(0)}={r_0\over \pi R^2}\, .
\eeq
It should be noticed that this is positive and finite in spite of the absence of the
excited matter of the system since $b_0=0$. On the other hand, this temperature is  
related to the energy and entropy at small $b_0$
as follows,
\beq\label{groundE}
  E(b_0) \simeq {3\pi^2\over 2}N^2 x^4{T^{(0)}_{\rm ent}}^4\, ,
\eeq
\beq
  S_{EE}\simeq{3\pi^2\over 2}N^2 x^4{T^{(0)}_{\rm ent}}^3\, , 
\eeq
where the terms are retained up to $O(x^4)$. These equations implies (i) that
the dynamical degrees of freedom (DOF) of the excitation due to $b_0$ decreases
to zero like $x^4$. (ii) Secondly, the necessary energy to excite one DOF from the ground state
of $b_0=0$ is $T_{ent}^{(0)}$. So there is an energy gap to make the lowest excitation.

This fact is the reflection of the confinement and the existence of the glueball
which has the lowest mass. Actually, 
the glueball mass $m_g$ for $J^{PC}=2^{++}$ state has been given as follows \cite{GNI14}
\beq\label{glueball-mass}
   m_g^2=4(n+1)(n+4){r_0^2\over R^4}=|\lambda| (n+1)(n+4)\, , \quad n=0,1,2, \dots\, .
\eeq
From this, the lowest glueball-mass is found as {$m_g^{(0)}=4r_0/R^2$.}
The relation between the mass $m_g$ and $T_{ent}^{(0)}$ is therefore rewritten as  
\beq\label{T-glueball}
   T_{ent}^{(0)}={m_g^{(0)}\over 4\pi }\, .
\eeq
{This implies that we need a small but finite energy to excite the vacuum
to the lowest excited state with glueballs of the lowest mass. This is independent
of $C$, and it is determined by the parameter $r_0$ or $\lambda$.}

\vspace{.3cm}
\noindent{\bf Trangent region, $b_0/r_0\sim 1$} ;

\begin{figure}[htbp]
\vspace{.3cm}
\begin{center}
\includegraphics[width=12cm]{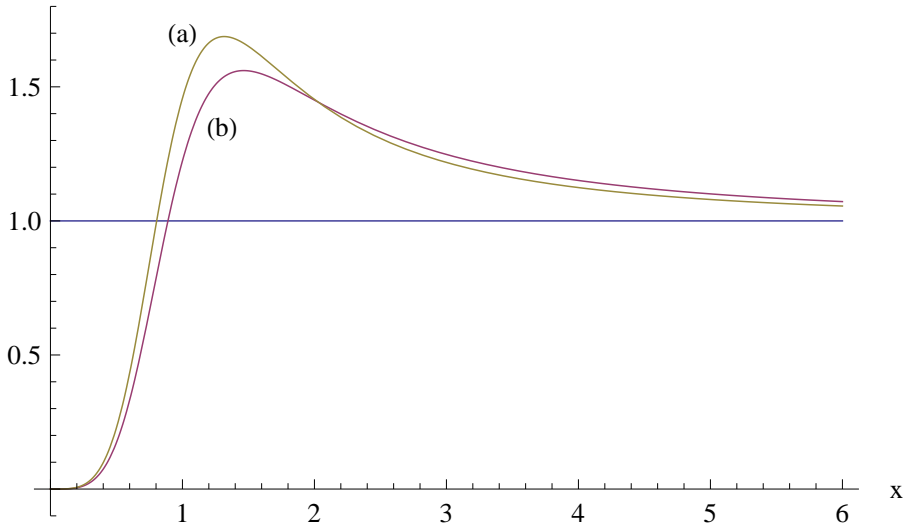}
\caption{(a) $f_U(x)$ and (b) {$f_S(x)$} are shown as the functions of $x=b_0/r_0$.
With decreasing $b_0$, the function $f_U$ ($f_S$) gradually increases from one to
its maximum $1.7 (1.6)$, which is realized at about $x=1.3 (1.4)$.
Then both $f_U$ and $f_S$ decrease rapidly to zero, which is realized at $b_0=0$.
}
\label{f-SU}
\end{center}
\end{figure}

The interesting point is seen in
the deviations from the high temperature limit, (\ref{largeb-S}) 
and (\ref{largeb-U}). 
They are expressed by the functions $f_U$ and $f_S$,
which are shown in the Fig. \ref{f-SU} as functions of $x=b_0/r_0$.

\begin{figure}[htbp]
\vspace{.3cm}
\begin{center}
\includegraphics[width=12cm]{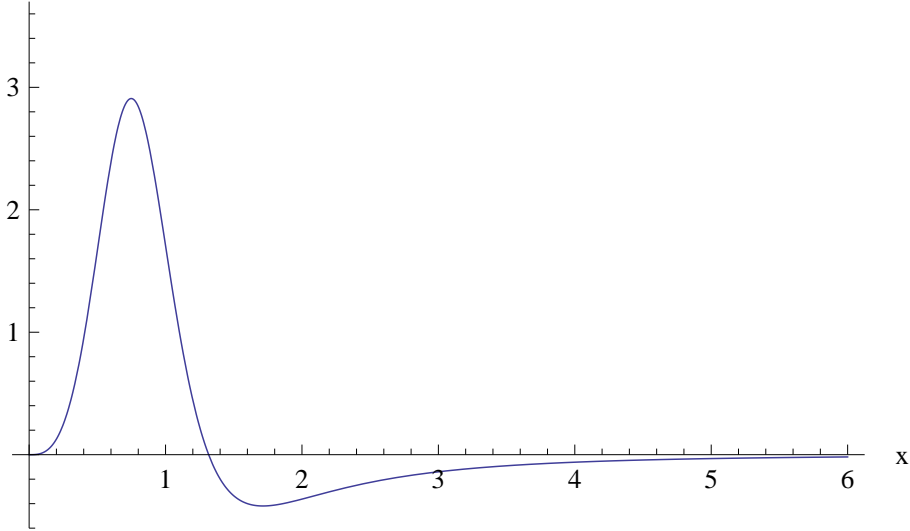}
\caption{(a) $\partial f_U(x)/\partial b_0$ is shown as the functions of $x=b_0/r_0$.
We find two extremum points, which would correspond to the two phase transitions
of the theory.
}
\label{df-SU}
\end{center}
\end{figure}

From the Fig.\ref{f-SU}, we can read the followings;
\begin{itemize}
 
\item For the range $0<x<0.4$, the values of $f_U$ and $f_S$ are almost zero. 
This is interpreted as 
the reflection of the confinement since the color degrees of freedom is suppressed
in this region probably to the one of $O(N^0)$. On the other hand, they increases rapidly
in the region of 
$0.4<b_0<1.0$ in spite of the fact that the theory is still in the confinement phase ($b_0<1$). 
This result could be related to the fact that the glueball mass is suppressed 
to smaller value when $b_0$ approaches to $b_0=r_0$, the critical point of (de) confinement
phase transition, as found in \cite{GNI14}. 
As a result, it would be possible to excite many higher order states of
glueballs since their mass spectrum would be given by the Eq. (\ref{glueball-mass})
with a small prefactor.

\item We should notice that 
this rapid variation of DOF has been also observed in the lattice simulation
of $SU(3)$ gauge theory near the (de) confinement transition temperature \cite{Baza}. 
In this case, however, the observed phenomenon is interpreted as 
the crossover. {Namely} it is not the first order transition. Therefore 
the maximum point of the
its increasing rate ($\partial f_U(x)/\partial b_0$) is identified
as the cross over point from the confinement
to the de-confinement. In this region, glueballs and color degrees of freedom coexist.

In the present case, however, this transition has been observed 
as the first order one \cite{GN13}, and the thermodynamic
property is examined by using the temperature which is defined by the Hawking temperature
$T_H$.
On the other hand, we are now considering the {extended} thermodynamics in terms of the
entanglement temperature $T_{ent}$. 

We could say that the order of the phase transition, which is observed
in the thermodynamics defined by $T_{ent}$, would be made milder than the one
given in the analysis in terms of the temperature $T_H$. Further, the critical
point given by the maximum of $\partial f_U(x)/\partial b_0$ is slightly smaller
than the actual transition point $b_0/r_0=1$.

\item Another point to be noticed is found by comparing our $f_U$ with the one
given in \cite{Baza}. Our $f_U$ has a maximum near $b_0/r_0=1$.
On the other hand, in the case of the $SU(3)$ lattice gauge simulation,
the corresponding factor $f_U$ increases monotonically without any such a maximum.

As shown in Fig.\ref{df-SU}, the existence of the maximum of $f_U$ implies the existence of
the minimum point of $\partial f_U(x)/\partial b_0$ also. This point also indicates the point
where $f_U$ changes rapidly, then it may
correspond to another phase transition point. It might be regarded
as the chiral transition point which has been found in our present model used here.
\end{itemize}

\section{Summary and Discussion}

In terms of entanglement entropy, we have examined the 
SYM theory, which is living in AdS$_4$ space-time.
The ground state of this theory is in the confinement phase, 
where we could observe the glueballs as excited modes of the theory. The mass
of the glueballs {is} expressed by the scale $r_0$ which characterizes AdS$_4$ curvature.

This theory can be extended to an excited state by adding 
extra parameter $C$ (or $b_0$) which is responsible for
the excitation of the SYM theory in the background determined by $r_0$. This yields
its energy density $\langle T_{00}\rangle \propto b_0^4$ in addition to the one of the
ground state composed of $r_0$. At enough large $b_0(>r_0)$, 
this excitation changes
the ground state from the confinement phase to 
{the de-confinement phase with a finite temperature}.

In the de-confinement phase ($r_0<b_0$), the temperature of the system is given
by the Hawking temperature $T_H(b_0,r_0)$, {which depends on $b_0$ and $r_0$}. {However,}  
$T_H(b_0,r_0)$ {disappears}
in the confinement phase $0<b_0\leq r_0$. 
So, in order to describe all region of the parameter $b_0$ as a 
thermodynamic phenomenon, we here introduced the entanglement temperature 
$T_{ent}$, {which is available in any phase. 
It} is derived by supposing the thermodynamic 
relation 
between the variations of
the energy density $\langle T_{00}\rangle$ and the entanglement
entroy $S_{EE}(b_0)$. Thus we could obtain
the temperature $T_{ent}$ which is useful {at any value} of $b_0$.

We used the approximate formula for $S_{EE}(b_0)$, {which is evaluated} 
in the limit of
large radius of the 3D hyperboloid, $p_0$. 
This approximation is reasonable to see the
thermodynamic properties 
since $S_{EE}(b_0)$ approaches to the {Bekenstein-Hawking} entropy at large scale.
The regularization for $S_{EE}(b_0)$ is performed by subtracting $S_{EE}(b_0=0)$ since
our interest is in the region of $0\leq b_0$. As for the ultraviolet divergences occuring near
the boundary, we could see the expected results in the logarithmic term.
We should notice that in the calculation of $S_{EE}(b_0)$ we could
not find the topological change of the minimal surface from the connected 
to the disconnected one when $p_0$ increases as found in the case of AdS soliton model \cite{KKM}.

Using the obtained $T_{ent}$,  {the two quantities},
$\langle T_{00}\rangle /T_{ent}^4$ and $S_{EE}/T_{ent}^3$, which
{correspond} to the effective dynamical degrees of freedom, are considered.
Then how these quantities are deviate from their high temperature limit
is studied for all region of $b_0$ including the phase transition point of the theory. 

We could find that the dynamical {degree} of freedom reduces to the very small number 
at small $b_0\sim 0$. This fact is interpreted as the reflection of the color confinement
since the color {degree} of freedom of the excitation vanishes. Secondly, we notice that 
there is a lower bound for the temperature $T_{ent}$, namely $T_{ent}\geq T_{ent}^{(0)}$.
The lower bound $T_{ent}^{(0)}$ is related to the glueball mass $m_g$  
as $T_{ent}^{(0)}=m_g/(4\pi)$.
These facts {indicate} that finite minimum energy is necessary for 
the excitation from the ground state 
$b_0=0$, and the excitation corresponds to the formation of glueballs with the lowest mass.
This is also the reflection of the confinement phase at small $T_{ent}$.

Further, we find from $\langle T_{00}\rangle /T_{ent}^4$
that the dynamical degrees of freedom increase rapidly near 
the transition point, from the confinement to the de-confinement phase. 
This phenomenon is similar to the cross over transition observed in the lattice
QCD with $SU(3)$ color symmetry {with flavor quarks}. 
So we could understand that the thermodynamic 
phenomenon with the entanglement temperature would reproduce milder order phase
transition than the case of Hawking temperature $T_H$. 

{Thirdly}, since $\langle T_{00}\rangle /T_{ent}^4$ has a maximum, it decreases after passing 
through this maximum and approaches to the expected high temperature limit.
So there is a second extremum for the derivative 
of the deviation of the freedom. This point also could be regarded as the phase transition
ponit as discussed in \cite {Baza}. In our model, this transition would be interpreted as
the chiral restoration point.
Therefore, we could say that 
we can see the expected phase transitions as the thermodynamic phenomenon
in terms of the entanglement temperature $T_{ent}$.

\vspace{.3cm}
\section*{Acknowledgments}
K. G thanks to M. Tachibana, F. Toyoda and I. Maruyama for useful discussion.
The work of M. Ishihara is supported by World Premier International Research Center Initiative WPI, MEXT, Japan. The work of M. I. is supported in part by the JSPS Grant-in-Aid for Scientific Research, Grant No. 15K20877.

\newpage



\def\theequation{A. \arabic{equation}}
\setcounter{equation}{0}

\appendix

\noindent{\bf\Large Appendix}

\section{Comment on the boundary metric}

We notice the coordinate of the boundary for (\ref{10dmetric-2}). It is given as
\beq\label{10dmetric-21}
ds^2_{4}=-dt^2+a_0^2(t)\gamma_{ij}(x)dx^idx^j . 
\eeq
\beq\label{AdS4-300} 
    \gamma_{ij}(x)=\delta_{ij}\left( 1+k{\bar{r}^2\over 4\bar{r_0}^2} \right)^{-2}\, , \quad 
    \bar{r}^2=\sum_{i=1}^3 (x^i)^2\, ,
\eeq
where $k=\pm 1,$ or $0$. $\bar{r_0}$ denotes the scale factor of three space. Here we set as
$k=-1$ as stated in the Sec. 3. So the space is opend. Further (\ref{10dmetric-21}) is rewritten
by (\ref{gp-0}) and (\ref{gp-1}) in terms of the polar coordinate. Although the radial coordinate 
$p$ in this metric is restricted as $p<2$, the volume of the space is infinite and then opened. This point 
becomes more clear
when we rewrite the metric as
 \beq\label{10dmetric-211}
ds^2_{4}=-dt^2+a_0^2(t)\left( {dq^2\over 1+q^2}+q^2d\Omega_{(2)}^2 \right). 
\eeq
where
\beq\label{AdS4-301} 
    q={p\over 1-p^2/4}\, .
\eeq
The new radial coordinate is then set in the range of $0<q<\infty$. For $p\sim 2(=p_0)$, 
$q$ approaches to $\infty$, then the small change of $p$ near $p_0$ corresponds to a
very large change of $q$. It is more convenient to use $p$ than $q$ to performing 
especially the numeerical annalysis at large scale region.
Due to this reason, we used $p$ rather than $q$.


\newpage

\newpage
\end{document}